# Product-State Manifolds for $M$ Quantum Systems with $N$ Levels using the Fano form and the Induced Euclidean Metric

**Fotios D. Oikonomou**[1]

[1] University of Patras, Department of Physics, Patras 26500, Greece;
Correspondence: pheconom@upatras.gr; Tel.: (+30) 6977 906792

**Abstract:** In quantum mechanics separable states can be characterized as convex combinations of product states whereas non-separable states exhibit entanglement. Quantum entanglement has played a pivotal role in both theoretical investigations and practical applications within quantum information science. In this study, we explore the connection between product states and geometric structures, specifically manifolds and their associated geometric properties such as the first fundamental form (metric). We focus on the manifolds formed by the product states of M systems of N levels, examining the induced metric derived from the Euclidean metric. For elementary cases we will compute the Levi-Civita connection, and, where computationally tractable, the scalar curvature.

**Keywords:** Separable states, product states, entangled states, manifolds, surfaces, geometry, metric, curvature, Levi-Civita connection.

## 1. Introduction

The concept of quantum entanglement was first introduced by Erwin Schrödinger in 1935, who referred to it as "Verschränkung"—a German term later translated as "entanglement"—during the formative development of quantum mechanics. Around the same period, Einstein, Podolsky, and Rosen (EPR) identified entanglement as a fundamental and paradoxical feature of quantum theory, emphasizing its implications in their seminal EPR paper [1,2]. They drew attention to the existence of global quantum states in composite systems that cannot be represented as elementary tensor products of their constituent subsystems' states. This phenomenon reveals a deep-seated structure in the statistical correlations between components of a quantum system.

Entanglement is now recognized as a central concept in quantum information theory and plays a pivotal role in modern developments such as quantum communication, quantum computing, and quantum cryptography [3–5]. In contrast to separable states, entangled states exhibit non-classical correlations that cannot be explained by any local hidden variable theory [6,7]. A thorough understanding of the geometric and algebraic structure of these states is essential for foundational quantum theory as well as for the advancement of practical quantum technologies.

Among multipartite quantum states, product states represent the elementaryst case and form a foundational subset within the broader quantum state space. From a mathematical perspective, this subset exhibits rich structure, particularly when analyzed through the lens of differential geometry. The space of pure quantum states is the complex projective space $\mathbb{CP}^{N^M-1}$, and product states form a submanifold within this space. As such, their geometry can naturally be studied using tools from Riemannian geometry. In particular, quantities such as the induced metric, Levi-Civita connection, and scalar curvature provide insights into how the geometry reflects properties like quantum separability.

In this work, we investigate the geometric structure of the manifold formed by the product states of M quantum systems, each possessing N levels. We employ the Euclidean metric to induce a Riemannian structure on the manifold of product states. This approach is consistent with prior studies that have analyzed quantum state spaces using metrics such as the Fubini–Study and Bures metrics [8–10]. In low-dimensional cases, we explicitly compute geometric quantities—such as the Levi-Civita connection and scalar curvature—which offer both local and global characterizations of the product state manifold.

The primary aim of this work is to bridge the domains of quantum information theory and differential geometry by exploring how the geometric structure of the product state manifold reflects properties related to separability and entanglement. This geometric framework may yield novel tools for distinguishing entangled states from separable ones, or at least provide a deeper understanding of the "shape" and structure of separable states within the total state space.

For a product state $\rho_{AB} = \rho_A \otimes \rho_B$, the individual subsystems $\rho_A$ and $\rho_B$ fully describe the composite system. Only in such cases does a complete description of the whole follow directly from the parts. Any state that is not a product state exhibits some form of correlation and is generally referred to as a correlated state. In quantum theory, correlations manifest in a hierarchy, giving rise to distinct physical properties at different levels. The elementaryst of these are classically correlated or separable states, whose density matrices can be written in the form:

$$\rho = \sum_i p_i \rho_{Ai} \otimes \rho_{Bi} \text{ with } 0 \leq p_i \leq 1 \text{ and } \sum_i p_i = 1 \quad (1)$$

That is, separable states are convex mixtures of product states. Although it may be known that a given density matrix represents a separable state, there exists no general algorithm for constructing a decomposition of the form in Eq. (1), and such decompositions are typically non-unique. Peres and the Horodecki family [11,12] provided important criteria for characterizing separability, particularly in systems with Hilbert space dimensions 2×2 and 2×3. States that cannot be decomposed as in Eq. (1) are termed entangled.

Geometry often plays a central role in the study of entanglement, offering vivid and often intuitive insights into complex quantum phenomena. In this context, the objective of the present study is to identify and characterize the manifolds corresponding to product states in a system with M subsystems of N levels. Since separable states can be expressed as convex combinations of product states, understanding the geometry of these manifolds is crucial in distinguishing them from entangled states.

To this end, we employ the Fano form of the density matrix to describe M-qudit systems [14]. In such systems, we consider the following classifications:
- Totally product states: such as $\rho_{A_1 \cdots A_M} = \rho_{A_1} \otimes \cdots \otimes \rho_{A_M}$.
- P-product states: $\rho_{G_1 \cdots G_P} = \rho_{G_1} \otimes \cdots \otimes \rho_{G_P}$ where each $G_i$ is a subset of subsystems that may still be entangled internally. Special cases are the biproduct states $\rho_{IJK} = \rho_{IJ} \otimes \rho_K$ where $(I, J, K)$ is a permutation of $(A, B, C)$, for systems of three subsystems
- Convex combinations of the above, and
- Genuinely entangled states.

Convex combinations of totally product states form the set of fully separable states, while combinations of P-product states give P-separable states. In particular, convex combinations of biproduct states yield biseparable states [14].

The structure of the paper is as follows. Section 2 provides a concise review of essential concepts from differential geometry that form the theoretical foundation for the subsequent analysis. In particular in 2.1 we have a short motivating elementary example and in 2.2 follows the geometrical background. In Section 3 we review the complex manifolds of M qudits. In Section 4, we present the main results of the study. Specifically, Section 4.1 discusses the Bloch vector representation for both qubits and qudits. Section 4.2 introduces the Fano form for systems consisting of M subsystems with N levels and formulates systems of equations that characterize P-product, biproduct and fully product states. Section 4.3 is devoted to solving these systems and identifying the corresponding manifolds. In Section 4.4, we derive the first fundamental form (i.e., the induced Riemannian metric) on these manifolds.

Section 5 presents several illustrative examples and special cases. In Section 5.1, we analyze the product state for two qubits. In 5.2 we study biproduct states in a system of three qubits and compute the scalar curvature of the associated manifold. In Section 5.3, we examine the geometry of totally product states in a three-qubit system. Finally, Section 6 concludes the paper with a discussion of the results and outlines possible directions for future research.

## 2. Preliminary issues (Geometry)

### 2.1. A motivating simple example

Let us consider an "elementary" two-dimensional surface in three-dimensional space. It is well-established that this surface can be characterized by a function $F: \mathbb{R}^2 \to \mathbb{R}^3$ such that $F: (u^0, u^1) \mapsto (u^0, u^1, f(u^0, u^1))$ if certain conditions were satisfied, where $f: \mathbb{R}^2 \to \mathbb{R}$ some suitable map. This surface is, in essence, the graph of the function $f$. Let's see a generalization of these notions below.

### 2.2. Manifolds and geometry

Let us consider the n-dimensional manifold $\mathbb{R}^n$ and the m-dimensional manifold $\mathbb{R}^m$ with $m < n$. A point in $\mathbb{R}^m$ will be denoted $(u^0, \ldots, u^{m-1})$ and in $\mathbb{R}^n$ $(x^0, \ldots, x^{n-1})$. Let $F: \mathbb{R}^m \to \mathbb{R}^n$ be a $C^\infty$ function. If $p \in \mathbb{R}^m$, let $T_p\mathbb{R}^m$, $T_{F(p)}\mathbb{R}^n$ be the corresponding tangent spaces on $p$ and $F(p)$ respectively. Then $F$ induces the differential map $F_*: T_p\mathbb{R}^m \to T_{F(p)}\mathbb{R}^n$ defined by the relation $F_*V[f] = V[f \circ F]$ where $V = \sum V^i \frac{\partial}{\partial u^i} \in T_p\mathbb{R}^m$ and $f$ a function on $\mathbb{R}^n$. The map $F$ is called an immersion of $\mathbb{R}^m$ into $\mathbb{R}^n$ if $F_*$ is an injection. If $F$ is an injection and an immersion, that is an embedding, then $S = F(\mathbb{R}^m)$ is a submanifold of $\mathbb{R}^n$ [15]. The function $F$ also induces the pullback function $F^*: T^*_{F(p)}\mathbb{R}^n \to T^*_p\mathbb{R}^m$, where $T^*_{F(p)}\mathbb{R}^n$ and $T^*_p\mathbb{R}^m$ are the dual spaces of $T_{F(p)}\mathbb{R}^n$ and $T_p\mathbb{R}^m$ respectively. If $\omega = \sum \omega_i dx^i \in T^*_{F(p)}\mathbb{R}^n$ a one-form, the pullback $F^*$ is defined by the relation $\langle F^*\omega, V \rangle = \langle \omega, F_*V \rangle$ where for $X = \sum X^i \frac{\partial}{\partial x^i}$ we suppose $\langle \omega, X \rangle = \sum \omega_i X^i$ (the usual inner product). The pullback $F^*$ naturally extends to tensors of type (0,2) i.e. $\omega = \sum \omega_{ij} dx^i \otimes dx^j$.

The manifold $\mathbb{R}^n$ admits a natural metric called "Euclidean" $h = \sum \delta_{\alpha\beta} dx^\alpha \otimes dx^\beta$ where $\alpha, \beta = 0, \ldots, n-1$, and

$$\delta_{\alpha\beta} = \begin{cases} 1 \text{ if } \alpha = \beta \\ 0 \text{ if } \alpha \neq \beta \end{cases} \tag{2}$$

is the Kronecker delta. This means that if we make an infinitesimal step into $\mathbb{R}^n$, that is $x^\alpha \to x^\alpha + dx^\alpha$, then the length of this step is $ds$ where $ds^2 = \sum \delta_{\alpha\beta} dx^\alpha \otimes dx^\beta$. If we were on a two-dimensional flat surface this would be identical to "Pythagorean" theorem. In $\mathbb{R}^m$ we have an induced metric $g = F^*h$ i.e.

$$g = \sum_{\mu,\nu=0}^{m-1} g_{\mu\nu} du^\mu \otimes du^\nu \tag{3}$$

with [10]

$$g_{\mu\nu} = \sum_{\alpha,\beta=0}^{n-1} \delta_{\alpha\beta} \frac{\partial F^\alpha}{\partial u^\mu} \frac{\partial F^\beta}{\partial u^\nu} \quad (4)$$

and $\mu, \nu = 0, \ldots, m-1$. To clarify this notion, the induced metric on a two-dimensional surface embedded in three-dimensional space is used to compute Euclidean distances on the surface. Correspondingly, if we make an infinitesimal step into $\mathbb{R}^m$ such that $u^\mu \to u^\mu + du^\mu$, then the length will be $ds$ with

$$ds^2 = \sum_{\mu,\nu=0}^{m-1} g_{\mu\nu} du^\mu \otimes du^\nu. \quad (5)$$

For the manifold $S$ we may compute the connection coefficients $\Gamma^\lambda{}_{\mu\nu}$ of the Levi-Civita connection

$$\Gamma^\lambda{}_{\mu\nu} = \frac{1}{2} \sum_{\kappa=0}^{m-1} g^{\lambda\kappa} \left( \frac{\partial g_{\nu\kappa}}{\partial u^\mu} + \frac{\partial g_{\mu\kappa}}{\partial u^\nu} - \frac{\partial g_{\mu\nu}}{\partial u^\kappa} \right) \quad (6)$$

and subsequently the Riemann curvature tensor

$$R^\kappa{}_{\lambda\mu\nu} = \frac{\partial \Gamma^\kappa{}_{\nu\lambda}}{\partial u^\mu} - \frac{\partial \Gamma^\kappa{}_{\mu\lambda}}{\partial u^\nu} + \sum_{\eta=0}^{m-1} \Gamma^\eta{}_{\nu\lambda} \Gamma^\kappa{}_{\mu\eta} - \sum_{\eta=0}^{m-1} \Gamma^\eta{}_{\mu\lambda} \Gamma^\kappa{}_{\nu\eta} \quad (7)$$

the Ricci tensor

$$R_{\mu\nu} = \sum_{\lambda=0}^{m-1} R^\lambda{}_{\mu\lambda\nu} \quad (8)$$

and the scalar curvature

$$Q = \sum_{\mu,\nu=0}^{m-1} g^{\mu\nu} R_{\mu\nu}. \quad (9)$$

The scalar curvature is an intrinsic measure of curvature at a point in space that depends only on the metric (not on how the space is embedded in a larger space).

## 3. The complex manifolds of $M$ qudits

Let us suppose we are given $M$ qudits on the Hilbert space $\mathcal{H}_{A_1} \otimes \cdots \otimes \mathcal{H}_{A_M}$. If they are described by the pure state

$$|\psi\rangle = \sum \psi_{i_1 \cdots i_M} |i_1\rangle_{A_1} \otimes \cdots \otimes |i_M\rangle_{A_M} \quad (10)$$

we have $N^M - 1$ complex degrees of freedom and the corresponding space is $\mathbb{CP}^{N^M-1}$. In case of a product pure state

$$|\psi\rangle = |\varphi\rangle_{A_1} \otimes \cdots \otimes |\varphi\rangle_{A_M} \quad (11)$$

we have $M(N-1)$ complex degrees of freedom and the corresponding space is

$$\underbrace{\mathbb{CP}^{N-1} \times \cdots \times \mathbb{CP}^{N-1}}_{M \text{ times}} \quad (12)$$

embedded in $\mathbb{CP}^{N^M-1}$ via the Segre embedding.

If the $M$ qudits are described by a density matrix $\rho$ the space of all mixed states is

$$\{\rho \in \mathbb{C}^{N^M \times N^M} \mid \rho^\dagger = \rho,\ \rho \geq 0,\ Tr\rho = 1\} \tag{13}$$

with real dimension $N^{2M} - 1$.

## 4. The manifolds of $M$ qudits using Fano form

*4.1. Bloch sphere expansion form of one qubit and generalization for a qudit*

Let $\rho_A$ be the density matrix of one qubit. Then we can expand it

$$\rho_A = \frac{1}{2}(I_2 + a_1\sigma^1 + a_2\sigma^2 + a_3\sigma^3) \tag{14}$$

where $I_2$ is the $2 \times 2$ identity matrix, $\sigma^i$, $i = 1,2,3$, are the Pauli matrices and $\vec{a} = (a_1, a_2, a_3)$ is the radius of the so-called Bloch sphere (Bloch vector), with $|\vec{a}| \leq 1$.

In case of a qudit $\rho_A$ the expansion takes the form

$$\rho_A = \frac{1}{N}\left(I_N + \sum_{i=1}^{N^2-1} a_i\sigma^i\right) \tag{15}$$

where $I_N$ is the $N \times N$ identity matrix, $\vec{a} = (a_1, \ldots, a_{N^2-1})$ the Bloch vector and $\{\sigma^i\}_{i=1}^{N^2-1}$ are the Hermitian, traceless generators of the Lie algebra $\mathfrak{su}(N)$ — the analogs of the Pauli matrices. That means that we have a base given by the "vectors" $(e^0, e^1, \ldots, e^{N^2-1}) = \frac{1}{N}(I_N, \sigma^1, \ldots, \sigma^{N^2-1})$. We choose the normalization such that $tr(e^i \cdot e^j) = \frac{1}{N}\delta_{ij}$ where $i,j = 0, \ldots, N^2 - 1$.

*4.2. Fano form of M qudits*

If $\rho_{AB}$ is a density matrix of two qubits, we have the Fano form expansion [13]. The Fano form is explicitly stated in [9] for a quantum system consisted from two subsystems A, B of N levels

$$\rho_{AB} = \frac{1}{N^2}\left[I_{N^2} + \sum_{i=1}^{N^2-1} \alpha_i\sigma^i \otimes I_N + \sum_{j=1}^{N^2-1} b_j I_N \otimes \sigma^j + \sum_{i,j=1}^{N^2-1} \beta_{ij}\sigma^i \otimes \sigma^j\right] \tag{16}$$

where $\vec{a} = (a_1, \ldots, a_{N^2-1})$ and $\vec{b} = (b_1, \ldots, b_{N^2-1})$ are the Bloch vectors of the partially reduced states and $\beta_{ij} \in \mathbb{R}$. Given the above definitions we can compactly write for a bipartite system

$$\rho_{A_1 A_2} = \sum_{i,j=0}^{N^2-1} d_{ij} e^i \otimes e^j \tag{17}$$

with $d_{ij} \in \mathbb{R}$, $d_{00} = 1$ and such that $\rho_{A_1 A_2}$ is a density matrix. For a quantum system consisting of three N-level subsystems

$$\rho_{A_1 A_2 A_3} = \sum_{i,j,k=0}^{N^2-1} d_{ijk} e^i \otimes e^j \otimes e^k \tag{18}$$

with $d_{ijk} \in \mathbb{R}$, $d_{000} = 1$ and such that $\rho_{A_1 A_2 A_3}$ is a density matrix. The generalization for M subsystems is obvious

$$\rho_{A_1 \cdots A_M} = \sum_{i_1, \ldots, i_M = 0}^{N^2-1} d_{i_1 \cdots i_M} e^{i_1} \otimes \cdots \otimes e^{i_M} \tag{19}$$

Again $d_{i_1\cdots i_M}\epsilon\mathbb{R}$, $d_{0\cdots 0}=1$ and such that $\rho_{A_1\cdots A_M}$ is a density matrix. Rewriting the expansion (15) in the form

$$\rho_{A_i} = \sum_{j=0}^{N^2-1} a_j^i e^j \qquad (20)$$

with $a_0^i = 1$ for all $i = 1, \ldots, M$, we have the following algebraic systems. In case of (totally) product states we have, respectively, for $\rho_{A_1 A_2} = \rho_{A_1} \otimes \rho_{A_2}$

$$a_i^1 a_j^2 = d_{ij} \qquad (21)$$

where $i, j = 0, \ldots, N^2 - 1$.

For $\rho_{A_1 A_2 A_3} = \rho_{A_1} \otimes \rho_{A_2} \otimes \rho_{A_3}$ we have

$$a_i^1 a_j^2 a_k^3 = d_{ijk} \qquad (22)$$

where $i, j, k = 0, \ldots, N^2 - 1$ and in general, for

$$\rho_{A_1\cdots A_M} = \rho_{A_1} \otimes \cdots \otimes \rho_{A_M} \qquad (23)$$

we have

$$a_{i_1}^1 a_{i_2}^2 \cdots a_{i_M}^M = d_{i_1 i_2 \ldots i_M} \qquad (24)$$

where $i_l = 0, \ldots, N^2 - 1$ and $l = 1, \ldots, M$.

In addition, for a system, say, of three subsystems there exist the so-called biproduct states, i.e. states of the form $\rho_{A_1 A_2 A_3} = \rho_{A_1 A_2} \otimes \rho_{A_3}$. Given the expansions (17), (18), (20), we have

$$d_{ij} a_k^3 = d_{ijk} \qquad (25)$$

for $i, j, k = 0, \ldots, N^2 - 1$ and similarly for the other cases $\rho_{A_1 A_3} \otimes \rho_{A_2}$ or $\rho_{A_2 A_3} \otimes \rho_{A_1}$. In general, we consider a partition $\{I_1, \ldots, I_P\}$ of the set of indices $I = \{i_1, i_2, \ldots, i_M\}$ where $I_l$, $l = 1, \ldots, P$, are disjoint subsets of this set ($\cup_{l=1}^P I_l = I$). The P-product (not totally) states are characterized by the algebraic system

$$d_{I_1} d_{I_2} \cdots d_{I_P} = d_{i_1 i_2 \ldots i_M} \qquad (26)$$

where $d_{I_l}$ has the indices belonging to $I_l$. This algebraic system has the same algebraic form with the system (24) of the totally product states, so, we will not study it separately except for a special case.

*4.3. Solving the system for a product state*

According to (23) and (24) the state $\rho_{A_1\cdots A_M}$ is a totally product state iff the algebraic system (24) with the unknows $a_{i_1}^1, a_{i_2}^2, \ldots, a_{i_M}^M$ has a solution. Special cases are the systems (25) and (26). The system (24) is a nonlinear system of $(N^2)^M - 1 = N^{2M} - 1$ equations with $(N^2 - 1)M$ unknows. So, it is an overdetermined system. If we find the conditions for this system to have a solution, these conditions will be what we want for the state $\rho_{A_1\cdots A_M}$ to be a product state.

If we set all $i_1, i_2, \ldots, i_M = 0$, except $i_l$ with $1 \leq l \leq M$, from (24) we take

$$\begin{aligned} a_{i_1}^1 a_{i_2}^2 \cdots a_{i_M}^M &= d_{i_1 i_2 \ldots i_M} \Rightarrow \\ a_0^1 a_0^2 \cdots a_{i_l}^l \cdots a_0^M &= d_{00\ldots i_l\ldots 0} \Leftrightarrow \\ 1 \cdot 1 \cdots a_{i_l}^l \cdots 1 &= d_{00\ldots i_l\ldots 0} \Leftrightarrow \\ a_{i_l}^l &= d_{00\ldots i_l\ldots 0} \end{aligned} \qquad (27)$$

So, the necessary and sufficient conditions for the system (24) to have a solution are

$$d_{i_1 0 \ldots 0} d_{0 i_2 \ldots 0} \cdots d_{00 \ldots i_M} = d_{i_1 i_2 \ldots i_M}. \tag{28}$$

Moreover, these are the conditions for the state $\rho_{A_1 \cdots A_M}$ to be a (totally) product state. Now we will consider the above equations (28) as the definition of a function $F$

$$F: \mathbb{R}^{(N^2-1)M} \to \mathbb{R}^{N^{2M}-1} \tag{29}$$

such that

$$F: (d_{10\ldots 0}, \ldots, d_{N^2-1,0\ldots 0}, d_{01\ldots 0}, \ldots, d_{0,N^2-1\ldots 0}, \ldots, d_{00\ldots 1}, \ldots, d_{00\ldots N^2-1}) \\ \mapsto (d_{i_1 i_2 \ldots i_M})_{\substack{i_1,i_2,\ldots,i_M=0 \\ \text{except all } 0}}^{N^2-1} \tag{30}$$

If we make the correspondence

$$(x^\alpha)_{\alpha=0}^{N^{2M}-2} \equiv (d_{i_1 i_2 \ldots i_M})_{\substack{i_1,i_2,\ldots,i_M=0 \\ \text{except all } 0}}^{N^2-1}$$

$$(u^\mu)_{\mu=0}^{(N^2-1)M-1} \tag{31}$$

$$\equiv (d_{10\ldots 0}, \ldots, d_{N^2-1,0\ldots 0}, d_{01\ldots 0}, \ldots, d_{0,N^2-1\ldots 0}, \ldots, d_{00\ldots 1}, \ldots, d_{00\ldots N^2-1})$$

then this function is of the form $F: (u^0, \ldots, u^{(N^2-1)M-1}) \mapsto (x^0, \ldots, x^{N^{2M}-2})$ and in particular

$$F: (u^0, \ldots, u^{(N^2-1)M-1}) \\ \mapsto (u^0, \ldots, u^{(N^2-1)M-1}, f^1(u^0, \ldots, u^{(N^2-1)M-1}), \ldots, f^{N^{2M}-1-(N^2-1)M}(u^0, \ldots, u^{(N^2-1)M-1})) \tag{32}$$

as in the elementary motivational example mentioned above.

This function is obviously $C^\infty$ and an injection (one to one). Indeed, if for two points $(u^0, \ldots, u^{(N^2-1)M-1}) \neq (u^{0\prime}, \ldots, u^{(N^2-1)M-1\prime})$ then $F(u^0, \ldots, u^{(N^2-1)M-1}) \neq F(u^{0\prime}, \ldots, u^{(N^2-1)M-1\prime})$ since the first $(N^2-1)M$ coordinates of the two images are these vectors and are different. The differential map $F_*$ is also an injection. To prove this, it is known that, if $W = F_* V$ then $W^\alpha = \sum_\mu \frac{\partial F^\alpha}{\partial u^\mu} V^\mu$. Given the form of $F$ above, it is obvious that the $rank\left(\left[\frac{\partial F^\alpha}{\partial u^\mu}\right]\right) = (N^2-1)M$, since one minor determinant of the rectangular matrix $\left[\frac{\partial F^\alpha}{\partial u^\mu}\right]$ is the determinant of the identity matrix and it is different from 0. So, $F_*$ is an injection.

Finally, given the above analysis, the image of $F$ is a submanifold of $\mathbb{R}^{N^{2M}-1}$.

*4.4. The induced metric*

We can find the induced metric of the embedded submanifold. We have

$$\frac{\partial d_{i_1 i_2 \ldots i_M}}{\partial d_{0 \ldots i_j \ldots 0}} = d_{i_1 0 \ldots 0} d_{0 i_2 \ldots 0} \cdots \hat{d}_{0 \ldots i_j' \ldots 0} \cdots d_{00 \ldots i_M} \delta_{i_j i_j'} (1 - \delta_{0 i_j'}) \tag{33}$$

where the quantity below the hut is omitted. Then the components of the metric are, according to (4),

$$\sum_{\substack{i_1,i_2,\ldots,i_M=0 \\ \text{except all } 0}}^{N^2-1} \frac{\partial d_{i_1 i_2 \ldots i_M}}{\partial d_{0 \ldots i_j \ldots 0}} \frac{\partial d_{i_1 i_2 \ldots i_M}}{\partial d_{0 \ldots i_l \ldots 0}} =$$

$$= \sum_{\substack{i_1,i_2,\ldots,i_M=0 \\ \text{except all } 0}}^{N^2-1} d_{i_1 0 \ldots 0} d_{0 i_2 \ldots 0} \cdots \hat{d}_{0 \ldots i_j' \ldots 0} \cdots d_{00 \ldots i_M} \delta_{i_j i_j'} \left(1 - \delta_{0 i_j'}\right) \times$$

$$d_{i_1 0 \ldots 0} d_{0 i_2 \ldots 0} \cdots \hat{d}_{0 \ldots i_l' \ldots 0} \cdots d_{00 \ldots i_M} \delta_{i_l i_l'} \left(1 - \delta_{0 i_l'}\right) \tag{34}$$

for $i_j, i_l = 1, \ldots, N^2 - 1$ and $j, l = 1, \ldots, M$.

## 5. Special cases

*5.1. The product state for two qubits*

Initially, we will consider, as a special case, two qubits $A_1, A_2$. The relations (17) and (20) now take the form

$$\rho_{A_1 A_2} = \sum_{i,j=0}^{3} d_{ij} e^i \otimes e^j \qquad (35)$$

$$\rho_{A_i} = \sum_{j=0}^{3} a_j^i e^j. \qquad (36)$$

We will have a product state $\rho_{A_1 A_2} = \rho_{A_1} \otimes \rho_{A_2}$, iff relation (21) holds

$$a_i^1 a_j^2 = d_{ij}. \qquad (37)$$

This system can be solved given that $d_{00} = a_0^1 = a_0^2 = 1$

$$\begin{aligned} a_i^1 &= d_{i0} \text{ for } i = 0,1,2,3 \\ a_j^2 &= d_{0j} \text{ for } j = 0,1,2,3. \end{aligned} \qquad (38)$$

Since the system (37) is overdetermined the following conditions must be fulfilled

$$d_{i0} d_{0j} = d_{ij} \text{ for } i, j = 0,1,2,3 \qquad (39)$$

which are the relations we are looking for. We will consider the function $J$

$$\begin{aligned} J &: (d_{01}, d_{02}, d_{03}, d_{10}, d_{20}, d_{30}) \\ &\mapsto (d_{01}, d_{02}, d_{03}, d_{10}, d_{11}, d_{12}, d_{13}, d_{20}, d_{21}, d_{22}, d_{23}, d_{30}, d_{31}, d_{32}, d_{33}) \end{aligned} \qquad (40)$$

defined by the relations (39). Function $J$ is a $C^\infty$ function from $\mathbb{R}^6$ to $\mathbb{R}^{15}$. So, we have a 6-dimensional (hyper)surface $S_1$ embedded in the 15-dimensional manifold $\mathbb{R}^{15}$. The (hyper)surface $S_1$ is the manifold of product states into the whole space of bipartite density matrices.

The induced metric according to eq. (4) is

$$\begin{aligned} g_{\mu\mu} &= \sum_{i=0}^{3} d_{i0}^2 \text{ for } \mu = 0,1,2 \\ g_{\mu\mu} &= \sum_{i=0}^{3} d_{0i}^2 \text{ for } \mu = 3,4,5 \\ g_{\mu\nu} &= d_{0,\nu+1} d_{\mu-2,0} \text{ for } \mu = 3,4,5, \ \nu = 0,1,2 \\ g_{\nu\mu} &= g_{\mu\nu} \end{aligned} \qquad (41)$$

The other components of $g$ are 0. Then we procced computing the connection coefficients of the Levi-Civita connection (eq. (6)) the Riemann and Ricci tensors and finally the scalar curvature. Using "mathematica" we have for the scalar curvature $Q$:

$$\begin{aligned} q_1 &= \sum_{i=1}^{3} d_{0i}^2 \\ q_2 &= \sum_{i=1}^{3} d_{i0}^2 \\ Q &= -\frac{2(3 + 3q_1 + 2q_2)(3 + 2q_1 + 3q_2)}{(1 + q_1)(1 + q_2)(1 + q_1 + q_2)} \end{aligned} \qquad (42)$$

It is obvious that

$$Q < 0 \qquad (43)$$

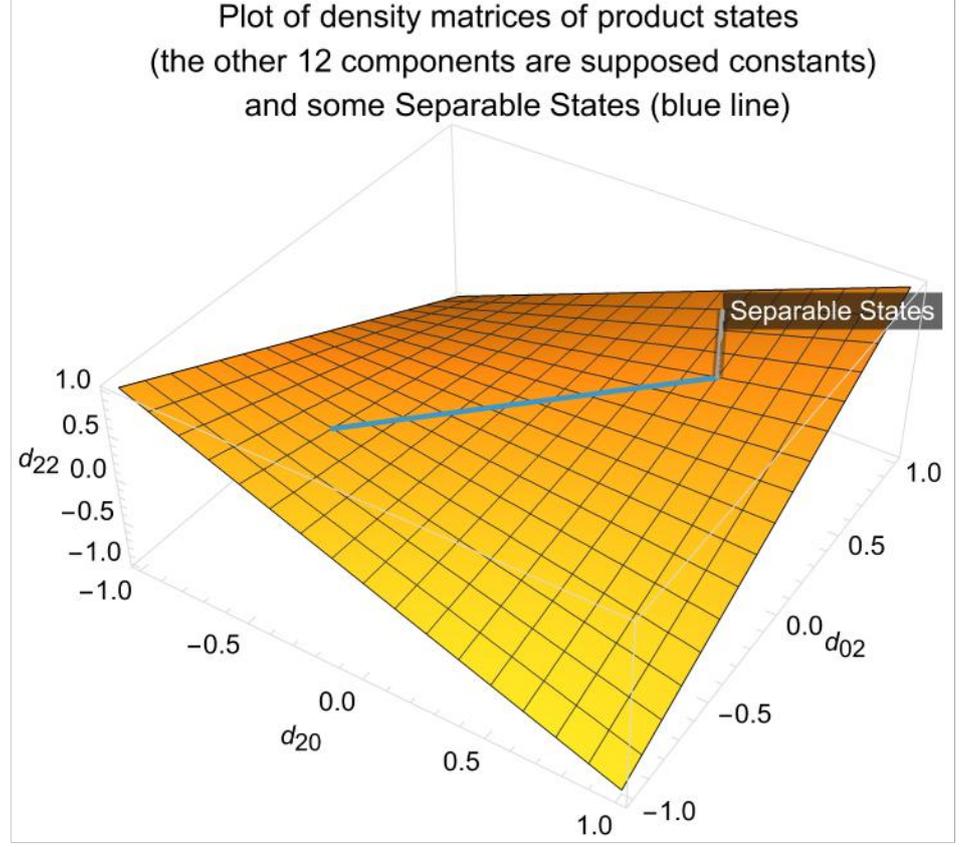

**Figure 1.** The function $d_{20}d_{02} = d_{22}$ is plotted to illustrate product states. The other components may be assumed constants satisfying (39).

*5.2. The biproduct state for three qubits*

Let's consider in detail the case of three qubits, that is, $M = 3$ and $N = 2$. The relations (17), (18) and (20) now take the form

$$\rho_{A_1 A_2} = \sum_{i,j=0}^{3} d_{ij} e^i \otimes e^j \quad (44)$$

$$\rho_{A_1 A_2 A_3} = \sum_{i,j,k=0}^{3} d_{ijk} e^i \otimes e^j \otimes e^k \quad (45)$$

$$\rho_{A_i} = \sum_{j=0}^{3} a_j^i e^j \quad (46)$$

We will have a biproduct state $\rho_{A_1 A_2 A_3} = \rho_{A_1 A_2} \otimes \rho_{A_3}$, iff relation (25) holds

$$d_{ij} a_k^3 = d_{ijk} \quad (47)$$

for $i, j, k = 0,1,2,3$.

This is can be easily solved given that $d_{00} = a_0^3 = d_{000} = 1$. It is obvious that its solution is

$$d_{ij} = d_{ij0} \text{ for } i,j = 0,1,2,3$$
$$a_k^3 = d_{00k} \text{ for } k = 0,1,2,3 \tag{48}$$

But, since the system is overdetermined, we have to impose the conditions

$$d_{00k} d_{ij0} = d_{ijk} \text{ for } i,j,k = 0,1,2,3. \tag{49}$$

Now, let $K$ be the function

$$K: (d_{00k}, d_{ij0})_{k=1, i,j=0 \text{ except both } 0}^{3, 3} \mapsto (d_{00k}, d_{ij0}, d_{ij1}, d_{ij2}, d_{ij3})_{k=1, i,j=0 \text{ except both } 0}^{3, 3} \tag{50}$$

defined by relations (49). To be precise the order of the coordinates of the domain of $K$ is

$(d_{001}, d_{002}, d_{003}, d_{010}, d_{020}, d_{030}, d_{100}, d_{110}, d_{120}, d_{130}, d_{200}, d_{210}, d_{220}, d_{230}, d_{300}, d_{310}, d_{320}, d_{330})$

and for the image

$(d_{001}, d_{002}, d_{003}, d_{010}, d_{020}, d_{030}, d_{100}, d_{110}, d_{120}, d_{130}, d_{200}, d_{210}, d_{220}, d_{230}, d_{300}, d_{310}, d_{320}, d_{330},$
$d_{011}, d_{012}, d_{013}, d_{021}, d_{022}, d_{023}, d_{031}, d_{032}, d_{033}, d_{101}, d_{102}, d_{103}, d_{111}, d_{112}, d_{113}, d_{121}, d_{122}, d_{123},$
$d_{131}, d_{132}, d_{133}, d_{201}, d_{202}, d_{203}, d_{211}, d_{212}, d_{213}, d_{221}, d_{222}, d_{223}, d_{231}, d_{232}, d_{233}, d_{301}, d_{302}, d_{303},$
$d_{311}, d_{312}, d_{313}, d_{321}, d_{322}, d_{323}, d_{331}, d_{332}, d_{333}).$

We notice that $K$ is a $C^\infty$ function from $\mathbb{R}^{18}$ to $\mathbb{R}^{63}$ since $3 + 4 \cdot 4 - 1 = 18$ and $3 + 4 \cdot (4 \cdot 4 - 1) = 63$ respectively. So, we have an 18-dimensional (hyper)surface $S_2$ embedded in the 63-dimensional manifold $\mathbb{R}^{63}$. The (hyper)surface $S_2$ is the manifold of biproduct states into the whole space of tripartite density matrices.

If $h$ is the Euclidean metric in $\mathbb{R}^{63}$ we can compute the induced metric $g = K^*h$. The result is (eq. (4))

$$g_{\mu\mu} = \sum_{i,j=0}^{3} d_{ij0}^2 \text{ for } \mu = 0,1,2$$
$$g_{\mu\mu} = \sum_{i=0}^{3} d_{00i}^2 \text{ for } \mu = 3,\ldots,17$$
$$g_{\mu\nu} = d_{00,\nu+1} m_{\mu-2} \text{ for } \mu = 3,\ldots,17, \nu = 0,1,2 \tag{51}$$
$$g_{\nu\mu} = g_{\mu\nu}$$

where $\vec{m} = (m_0, \ldots, m_{15}) = (d_{000}, d_{010}, \ldots, d_{320}, d_{330})$. The other components of $g$ are 0.

Then we proceed computing the connection coefficients of the Levi-Civita connection (eq. (6)) the Riemann and Ricci tensors and finally the scalar curvature. Using "mathematica" we have for the scalar curvature $Q$:

$$q_1 = \sum_{i=1}^{3} d_{00i}^2$$
$$q_2 = 1 + q_1 + \sum_{i=1}^{3} d_{0i0}^2 + \sum_{i=1}^{3} \sum_{j=0}^{3} d_{ij0}^2 \tag{52}$$
$$Q = -\frac{2(q_1 - 3q_2)(1 + q_1 + 14q_2)}{(1 + q_1)(q_1 - q_2)q_2^2}.$$

Since $q_1 - q_2 < 0$ and $q_1 < 3q_2$. we have

$$Q < 0. \tag{53}$$

*5.3. The product state for three qubits*

In the case of the product states of three qubits $\rho_{A_1A_2A_3} = \rho_{A_1} \otimes \rho_{A_2} \otimes \rho_{A_3}$ we have the system (22) $a_i^1 a_j^2 a_k^3 = d_{ijk}$. We will solve is this system for $a_0^1 = a_0^2 = a_0^3 = d_{000} = 1$. It is obvious that its solution is

$$a_i^1 = d_{i00} \text{ for } i = 0,1,2,3$$
$$a_j^2 = d_{0j0} \text{ for } j = 0,1,2,3 \quad (54)$$
$$a_k^3 = d_{00k} \text{ for } k = 0,1,2,3$$

with the following conditions to be fulfilled

$$d_{i00} d_{0j0} d_{00k} = d_{ijk} \text{ for } i,j,k = 0,1,2,3 \quad (55)$$

Correspondingly $L$ will be the function

$$L: (d_{i00}, d_{0j0}, d_{00k})_{i,j,k=1}^{3} \mapsto (d_{ijk})^3_{\substack{i,j,k=0 \\ \text{except all } 0}} \quad (56)$$

defined by relations (55).

To be precise again the order of the coordinates of the domain of $L$ is

$$(d_{100}, d_{200}, d_{300}, d_{010}, d_{020}, d_{030}, d_{001}, d_{002}, d_{003})$$

and for the image

$$(d_{001}, d_{002}, d_{003}, d_{010}, d_{020}, d_{030}, d_{100}, d_{110}, d_{120}, d_{130}, d_{200}, d_{210}, d_{220}, d_{230}, d_{300}, d_{310}, d_{320}, d_{330},$$
$$d_{011}, d_{012}, d_{013}, d_{021}, d_{022}, d_{023}, d_{031}, d_{032}, d_{033}, d_{101}, d_{102}, d_{103}, d_{111}, d_{112}, d_{113}, d_{121}, d_{122}, d_{123},$$
$$d_{131}, d_{132}, d_{133}, d_{201}, d_{202}, d_{203}, d_{211}, d_{212}, d_{213}, d_{221}, d_{222}, d_{223}, d_{231}, d_{232}, d_{233}, d_{301}, d_{302}, d_{303},$$
$$d_{311}, d_{312}, d_{313}, d_{321}, d_{322}, d_{323}, d_{331}, d_{332}, d_{333})$$

the same as the coordinates in case of $K$.

We notice that $L$ is a function from $\mathbb{R}^9$ to $\mathbb{R}^{63}$ since $3 \cdot 3 = 9$ and $4^3 - 1 = 63$. So, we have a 9-dimensional (hyper)surface $S_3$ embedded in the 63-dimensional manifold $\mathbb{R}^{63}$. The (hyper)surface $S_3$ is the manifold of totally product states into the whole space of tripartite density matrices.

As before if $h$ is the Euclidean metric in $\mathbb{R}^{63}$ we can compute the induced metric $g = L^*h$. If

$$r_1 = d_{000}^2 + d_{001}^2 + d_{002}^2 + d_{003}^2$$
$$r_2 = d_{000}^2 + d_{010}^2 + d_{020}^2 + d_{030}^2 \quad (57)$$
$$r_3 = d_{000}^2 + d_{100}^2 + d_{200}^2 + d_{300}^2$$

the result is (eq. (4))

$$g_{\mu\mu} = r_1 r_2 \text{ for } \mu = 0,1,2$$
$$g_{\mu\mu} = r_1 r_3 \text{ for } \mu = 3,4,5$$
$$g_{\mu\mu} = r_2 r_3 \text{ for } \mu = 6,7,8$$
$$g_{\mu\nu} = r_1 d_{0,\mu-2,0} d_{\nu+1,0,0} \text{ for } \mu = 3,4,5, \nu = 0,1,2 \quad (58)$$
$$g_{\mu\nu} = r_2 d_{0,0,\mu-5} d_{\nu+1,0,0} \text{ for } \mu = 6,7,8, \nu = 0,1,2$$
$$g_{\mu\nu} = r_3 d_{0,0,\mu-5} d_{0,\nu-2,0} \text{ for } \mu = 6,7,8, \nu = 3,4,5$$
$$g_{\nu\mu} = g_{\mu\nu}.$$

The other components of $g$ are 0. We may compute the scalar curvature with "mathematica", but the expression is too big to be written here.

## 6. Discussion

The results obtained in this study highlight the potential of differential geometry as a powerful framework for analyzing the structure of product-state manifolds in multipartite quantum systems. By explicitly constructing the manifolds associated with both biproduct and totally product states, and by deriving their induced metrics from the Euclidean metric, we have demonstrated how geometric invariants such as curvature can encode information about quantum separability.

One of the key observations is that the manifolds of product states possess a well-defined Riemannian structure, whose scalar curvature can, in principle, serve as a quantitative descriptor of their local and global geometry. In the product state of two-qubit system and in the biproduct case for three-qubit systems, our explicit calculation of the scalar curvature revealed a consistently negative value, indicating a locally hyperbolic character of the manifold. Such curvature signatures could provide geometric criteria for distinguishing different classes of separability, complementing existing algebraic and entanglement-based measures.

The formalism developed here is general and extendable. While the two-qubit and three-qubit cases serve as a tractable example for explicit computation, the same methodology applies to systems with higher dimensions (N>2) and a larger number of subsystems (M>3). The primary limitation is computational complexity, as the dimensionality of the embedding space and the number of parameters grow rapidly with M and N. Nonetheless, symbolic and numerical computational tools, such as those employed here, can make such generalizations feasible.

Beyond pure mathematical interest, this geometric viewpoint could have practical implications. For instance, in quantum information processing, the ability to map separable and entangled states to distinct geometric regions might inform the design of quantum algorithms or error-correcting codes that exploit separability constraints. Moreover, since the induced metric depends on the choice of the underlying metric on the ambient state space, future work could compare the Euclidean-induced metric with other natural choices, such as the Fubini–Study or Bures metrics, to investigate whether curvature-based distinctions are metric-independent or metric-specific.

Another promising direction is to explore whether the geodesic structure of these manifolds correlates with physical processes such as local unitary evolution, separable operations, or decoherence. Understanding how product-state manifolds are embedded within the full state space may also shed light on the geometry of entanglement witnesses and the boundaries between separable and entangled regions.

In summary, this study provides a proof-of-concept for a systematic geometric characterization of product states in multipartite quantum systems. The approach merges concepts from quantum information theory with the formal machinery of differential geometry, potentially opening avenues for new analytical tools and classification schemes. While much remains to be explored—particularly in higher-dimensional and mixed-state settings—the present work lays a foundation for a richer understanding of the interplay between geometry and quantum separability.

**Acknowledgments**: The author wishes to express sincere gratitude to Emeritus Professor D. P. K. Ghikas for his valuable indications.

**Conflicts of Interest:** The author declare no conflicts of interest.

**Data availability statement:** No datasets were generated or analyzed during the current study.